\documentclass[preprintnumbers,showkeys,superscriptaddress,showpacs,
nofootinbib,byrevtex,fleqn]{revtex4}   
\usepackage{amsmath,amsfonts,amssymb,amscd,amsxtra,amsthm}
\usepackage{graphicx}
\usepackage{bm}
\usepackage{epstopdf}
\begin{document}  
\preprint{INHA-NTG-10/2010}
\title{Spin structure of the pion from the instanton vacuum}      
\author{Seung-il Nam}
\email[E-mail: ]{sinam@kau.ac.kr}
\affiliation{Research Institute of Basic Sciences, Korea Aerospace
University, Goyang, 412-791, Korea}  
\author{Hyun-Chul Kim}
\email[E-mail: ]{hchkim@inha.ac.kr}
\affiliation{Department of Physics, Inha University, Incheon 402-751,
Korea}  
\date{\today}
\begin{abstract}
We investigate the spin structure of the pion within the framework of
the nonlocal chiral quark model from the instanton vacuum. We first
evaluate the tensor form factors of the pion for the first and second
moment $(n=1,2)$ and compare it with the lattice data. Combining the
tensor form factor of the pion with the  electromagnetic one, we
determine the impact-parameter dependent probability density of
transversely polarized quarks inside the pion. It turns out that the
present numerical results for the tensor form factor as well as those
for the probability density are in good agreement with the lattice
data. We also discuss the distortion of the spatial distribution of
the quarks in the transverse plane inside the pion. 
\end{abstract} 
\maketitle
\section{Introduction}
The transversity of hadrons has been one of the most
important issues  well over decades (see a recent
review~\cite{Barone:2001sp}), since it allows one to get access to
their spin structures. It is pertinent to the tensor current of
hadrons and is very difficult to be measured experimentally, because
there is no direct probe to measure it. Only very recently, it was
suggested that the transverse spin asymmetry $A_{TT}$ in Drell-Yan
processes in $p\overline{p}$ reactions    
\cite{Efremov:2004,Anselimo:2004,PAX:2005,Pasquini:2006} as well as  
the azimuthal single spin asymmetry in semi-inclusive deep inelastic
scattering (SIDIS) \cite{Anselmino:tensorcharge} can be used to obtain 
information on the transversity of the nucleon. 
Though it is even more difficult to measure the transversity of the
pion experimentally, it is still of great significance to understand
it, since it provides the internal spin structure due to quarks,
i.e. it accommodates a novel concept called \textit{a hadron
  tomography}. The first result of the pion
transversity on lattice has been reported by the QCDSF/UKQCD  
Collaborations~\cite{Brommel:2007xd}. They also presented the
probability density of the polarized quarks inside the 
pion, combining the electromagnetic form factor of the
pion~\cite{Brommel:2006ww} with its tensor form factor. It was shown
in Ref.~\cite{Brommel:2007xd} that when the quarks are transversely
polarized, their spatial distribution is strongly distorted. This
first result in the lattice QCD has triggered 
several subsequent theoretical
works~\cite{Frederico:2009fk,Gamberg:2009uk,Broniowski:2010nt}.  
In Ref.~\cite{Broniowski:2010nt}, the tensor form factors of the pion
have been studied within the local and nonlocal Nambu-Jona-Lasinio
(NJL) model~\cite{Broniowski:2010nt}, a direct comparison with  
the lattice results being emphasized. In doing so, they employed a
larger value of the pion mass, i.e. $m_\pi = 600$ MeV such that the
results can be confronted with the lattice data. They also considered
the case of the chiral limit. 

In the present work, we first want to investigate the pion tensor form 
factor in the space-like momentum transfer region ($0\le
Q^2\le 1$ GeV), based on the low-energy effective chiral action
(E$\chi$A) from the instanton vacuum~\cite{Diakonov:1985eg}.  
Combining the result of the tensor form factor with the
electromagnetic one of the pion which was already studied in
Ref.~\cite{Nam:2007gf}  within the same framework, we then derive the 
probability density of transversely polarized quarks inside the pion. 
Since the instanton vacuum realizes the spontaneous chiral symmetry
breaking (S$\chi$SB) naturally via  
quark zero modes, it may provide a good framework to study properties 
of the pion such as the electromagnetic and tensor form factors.
Moreover, an important merit of this approach lies in the fact that
there are only two parameters, that is, the average (anti)instanton size
$\bar{\rho}\approx1/3$ fm and average
inter-instanton distance $\bar{R}\approx1$ fm.  The
normalization point is given by the average size of
instantons and is approximately equal to $\rho^{-1}\approx 0.6$ GeV. 
The values of the $\bar{\rho}$ and $\bar{R}$ were 
estimated many years ago phenomenologically in
Ref.~\cite{Shuryak:1981ff} as well as theoretically in 
Ref.~\cite{Diakonov:1983hh,Diakonov:2002fq,Schafer:1996wv}. The
instanton framework has been proved to be reliable in reproducing
experimental data especially for the meson sector, such as the meson
distribution amplitudes~\cite{Nam:2006sx,Nam:2006mb,Nam:2006au}, 
semileptonic decays~\cite{Nam:2007fx}, and etc. Furthermore, this
approach was supported by various lattice simulations of the QCD 
vacuum~\cite{Chu:vi,Negele:1998ev,DeGrand:2001tm}.  The quark
propagator from the instanton vacuum ~\cite{Diakonov:1983hh} is in a
remarkable agreement with lattice
calculations~\cite{Faccioli:2003qz,Bowman:2004xi}.  Finally the
nonlocal chiral quark model from the instanton vacuum has a practical
virtue, since it does not have any adjustable parameter once the
above-mentioned two parameters $\bar{\rho}$ and $\bar{R}$ are
determined.   

We organize the present work as follows: In Section II, we briefly 
explain the definitions of the probability densities of the transversely
polarized quarks, which are expressed  in terms of generalized form
factors of the pion.  In Section III, we show how to calculate the
tensor form factors within the nonlocal chiral quark model from the
instanton vacuum.  In Section IV, the numerical 
results are discussed and compared with those in lattice QCD.  The
final Section is devoted to summarize the present work, to draw
conclusions, and to give outlook.  

\section{Generalized form factors of the pion}
In this Section, we define the probability density of 
transversely polarized quarks inside the pion. For definiteness, we
choose the positively charged pion $\pi^{+}$ from now on. The
probability density of transversely polarized quarks for the $n$ th
moment of the probability density is given as  
\begin{equation}
\label{eq:DENSITY}
\rho_{n}(b_{\perp},s_{\perp})=\int^{1}_{-1}dx\,x^{n-1}
\rho(x,b_{\perp},s_{\perp})=
\frac{1}{2}\left[A _{n0}(b^{2}_{\perp})
-\frac{s^{i}_{\perp}\epsilon^{ij}b^{j}_{\perp}}{m_{\pi}}
\frac{\partial B _{n0}(b^{2}_{\perp})}{\partial b^{2}_{\perp}}\right],
\end{equation}
where $b_{\perp}$ denote the impact parameter that measures the
distance from the center of momentum of the pion to the quark in the
transverse plane to its motion.  The $s_\perp$ stands for the fixed
transverse spin of the quark. For simplicity, we choose the $z$
direction for the quark longitudinal momentum. The $x$ indicates the
momentum fraction possessed by the quark inside the pion. The
$A_{n0}(b_\perp^2)$ and $B_{n0}(b_\perp^2)$ are called the generalized
form factors (GFFs). In fact, the GFFs are just the moments of the
generalized parton distributions (GPDs) for the unpolarized and
transversely polarized pions, respectively:   
\begin{equation}
\label{eq:MOMENTS}
\int^{1}_{-1}dx\,x^{n-1}H (x,\xi=0,b^{2}_{\perp})
=A _{n0}(b^{2}_{\perp}),\,\,\,\,
\int^{1}_{-1}dx\,x^{n-1}E(x,\xi=0,b^{2}_{\perp})
=B _{n0}(b^{2}_{\perp}).
\end{equation}
For the first moment, the GFFs $A_{10}$ and $B_{10}$ are
identified with the electromagnetic and tensor form factors of the pion,
respectively. Previously, we have studied $A_{10}(q^2)$ in the
momentum space within the nonlocal chiral quark model (NL$\chi$QM)
from the instanton vacuum~\cite{Nam:2007gf}, resulting in a good
agreement with the experimental data. Hence, we can readily calculate 
$A_{10}(b_{\perp})$, using the results of Ref.~\cite{Nam:2007gf}.
Thus, we will concentrate on calculating the tensor form factors
$B_{10,20}$ within the same framework, and they can be written in a
general form as follows: 
\begin{equation}
\label{eq:BG}
\langle\pi^+(p_f)|\mathcal{O}^{\mu\nu\mu_1\cdots\mu_{n-1}}_T|\pi^+(p_i)\rangle
=\mathcal{AS}\left[\frac{(p^{\mu}q^{\nu}-q^{\mu}p^{\nu})}{m_{\pi}}
\sum^{n-1}_{i=\mathrm{even}}q^{\mu_1}\cdots q^{\mu_i}
p^{\mu_{i+1}}\cdots p^{\mu_{n-1}}B_{ni}(Q^2)\right],
\end{equation}
where $p_{i}$ and $p_{f}$ stand for the initial and final on-shell
momenta of  the pion, respectively. We also use notations
$p=(p_{f}+p_{i})/2$ and $q=p_{f}-p_{i}$. The tensor operator also can
be given as: 
\begin{equation}
\label{eq:OP}
\mathcal{O}^{\mu\nu\mu_1\cdots\mu_{n-1}}_T=
\mathcal{AS}
\left[q^\dagger\sigma^{\mu\nu}(i\tensor{D}^{\mu_1})
\cdots(i\tensor{D}^{\mu_{n-1}})q \right].
\end{equation}
The $\mathcal{A}$ and $\mathcal{S}$ denote the anti-symmetrization in
$(\mu,\nu)$ and symmetrization in $(\nu,\cdots,\mu_{n-1})$ with the
trace terms subtracted in all the indices. Taking into account
Eqs.~(\ref{eq:BG}) and (\ref{eq:OP}), we can define the
tensor form factors $B_{10}$ and $B_{20}$ of the pion 
in momentum space as the matrix elements of the tensor current, 
using the auxiliary-vector method as in Ref.~\cite{Diehl:2010ru}:  
\begin{eqnarray}
\label{eq:TENSOR}
&&\langle \pi^{+}(p_{f})|q^{\dagger}(0)\sigma_{ab}q(0)
|\pi^{+}(p_{i})\rangle =
\left[(p_i\cdot a)(p_f\cdot b)-(p_i\cdot b)(p_f\cdot a)\right]
\frac{B_{10}(Q^{2})}{m_\pi},
\cr
&&\langle \pi^{+}(p_{f})|q^{\dagger}(0)
\sigma_{ab}(i\tensor{D}\cdot a)q(0)
|\pi^{+}(p_{i})\rangle =
\left\{(p\cdot a)[(p_i\cdot a)(p_f\cdot b)-(p_i\cdot b)(p_f\cdot a)]
\right\} \frac{B_{20}(Q^{2})}{m_\pi},
\end{eqnarray}
where the vectors satisfy the conditions, i.e. $a^2=0$, $a\cdot b=0$
and $b^2\ne0$, and we have used a notation
$\sigma_{ab}\equiv\sigma_{\mu\nu}a^{\mu}b^{\nu}$. Due to this
auxiliary-vector method, one can eliminate the trace-term
subtractions. We also introduce a notation $i\tensor{D}_\mu\equiv 
(i\roarrow{D}_\mu-i\loarrow{D}_\mu)/2$, where $D_\mu$ indicates the
SU($N_c$) covariant derivative. Since we are interested in the spatial
distribution of the transversely polarized quark inside the pion, we
need to consider the Fourier transform of the form factors:   
\begin{equation}
\label{eq:FT}
F(b^{2}_{\perp})=\frac{1}{(2\pi)^2}\int d^{2} q_{\perp}
e^{-ib_{\perp}\cdot q_{\perp}}F(q^{2}_{\perp})
=\frac{1}{2\pi}\int^\infty_0 Q dQJ_0(bQ)F(Q^{2}),
\end{equation}
where $F=(A_{10},B_{10})$ designates the generic pion form
factor. The magnitudes of the transverse momentum and impact parameter
are expressed as $|\bm q_\perp|\equiv Q$ and $|\bm
b_\perp|\equiv b$. Similarly, the Fourier transform of the derivative
of the GFF with respect to $b^{2}_{\perp}$ can be evaluated as: 
\begin{equation}
\label{eq:DFT}
\frac{\partial F(b^{2}_{\perp})}{\partial b^{2}_{\perp}}
\equiv F'(b^{2}_{\perp})
=-\frac{1}{4\pi b}
\int^{\infty}_{0} Q^2dQJ_1(bQ)F(Q^{2}).
\end{equation}
The $J_0$ and $J_1$ in Eqs.(\ref{eq:FT},\ref{eq:DFT}) denote the
Bessel functions of order $0$ and $1$, respectively. 
According to the definitions for the relevant vectors $q_{\perp}$ and
$b_{\perp}$, the probability density in Eq.~(\ref{eq:DENSITY}) reads
as follows:  
\begin{equation}
\label{eq:DENSITY2}
\rho_{1}\left(b_{\perp},s_{x}=\pm 1\right)=
\frac{1}{2}\left[A _{10}(b^{2})
\mp\frac{b\sin\theta_\perp}{m_{\pi}}B' _{10}(b^{2})\right],
\end{equation}
where the spin of the quark inside the pion is quantized along the $x$
axis, $s_{\perp}=(\pm1,0)$.   

\section{Nonlocal chiral quark model from the 
instanton vacuum} 
We now briefly explain the NL$\chi$QM from the instanton 
vacuum~\cite{Diakonov:1995qy} and derive the GFFs of the
pion. Considering first the dilute instanton liquid, characterized by
the average (anti)instanton size $\bar{\rho}\approx1/3$ fm and average 
inter-instanton distance $\bar{R}\approx1$ fm with the small packing
parameter $\pi\bar{\rho}^4/\bar{R}^4\approx 0.1$, we are able to
average the fermionic determinant over collective coordinates of
instantons with fermionic quasi-particles, i.e. the constituent quarks
introduced. The averaged determinant is reduced to the light-quark
partition function that can be given as a functional of the tensor
field in the present case.  Having bosonized and integrated it over
the quark fields, we obtain the following effective nonlocal
chiral action in the large $N_c$ limit in Euclidean space: 
\begin{equation}
\label{eq:ACTION1}
\mathcal{S}_{\mathrm{eff}}[m,\pi]=-\mathrm{Sp}
\ln\left[i\rlap{/}{\partial}+im+i\sqrt{M(i\partial)}U^{\gamma_5}(\phi)
\sqrt{M(i\partial)}+\sigma\cdot T \right],
\end{equation}
where $m$, $\pi$, and $\mathrm{Sp}$ indicate the current quark mass,
the Nambu-Goldstone (NG) boson field, and the functional
trace over all relevant spaces, respectively. In the numerical
calculations, we will choose $m\sim m_{u}\sim m_{d}\approx5$ MeV,
taking into account isospin symmetry. The $M(i\partial)$ stands for 
the momentum-dependent effective quark mass, generated from the
nontrivial quark-(anti)instanton
interactions~\cite{Diakonov:1985eg}. Although  
its analytical form is in general given by the modified Bessel
functions, we will make use of its parametrization for numerical
convenience:    
\begin{equation}
\label{eq:EFFM}
M(i\partial)=M_{0}
\left(\frac{2}{2+\bar{\rho}^{2}\partial^{2}} \right)^{2},
\end{equation}
where $M_{0}$ indicates the constituent quark mass, which can be
determined self-consistently by solving the gap equation of the
present framework, resulting in $M_{0}=350$
MeV~\cite{Diakonov:1985eg}. The NG boson field is represented in a
nonlinear form as~\cite{Diakonov:1995qy}:    
\begin{equation}
\label{eq:CHIRALFIELD}
U^{\gamma_5}(\phi)=
\exp\left[\frac{i\gamma_{5}(\bm{\tau}\cdot\bm{\phi})}{F_{\phi}}\right]
=1+\frac{i\gamma_{5}(\bm{\tau}\cdot\bm{\phi})}{F_{\phi}}
-\frac{(\bm{\tau}\cdot\bm{\phi})^{2}}{2F^{2}_{\phi}}+\cdots,
\end{equation}
where $\phi^{a}$ is the SU(2) multiplet, defined as
\begin{equation}
\label{eq:PHI}
\bm{\tau}\cdot\bm{\phi}=\left(
\begin{array}{cc}
\frac{\pi^{0}}{\sqrt{2}}&\pi^{+}\\
\pi^{-}&\frac{\pi^{0}}{\sqrt{2}}\\
\end{array}
 \right).
\end{equation}
The $F_{\phi}$ denotes the weak-decay constant for NG bosons, whose 
empirical value is $93.2$ MeV for the pion for instance. The last term
in Eq.~(\ref{eq:ACTION1}) denotes $\sigma\cdot T=\sigma_{\mu\nu}
T_{\mu\nu}$, where $\sigma_{\mu\nu}=i[\gamma_\mu,\,\gamma_\nu]/2$ and
$T_{\mu\nu}$ represents the external tensor source field. 

\begin{figure}[h]
\includegraphics[width=10cm]{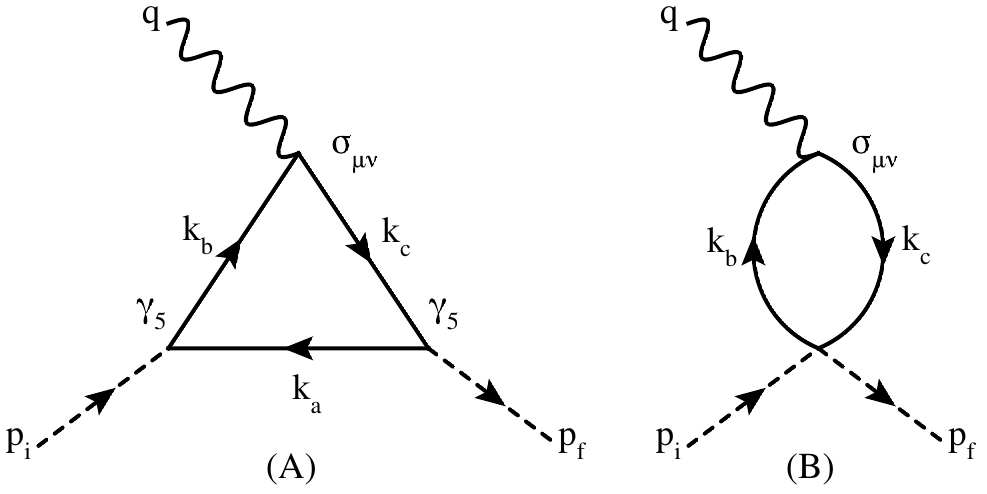}
\caption{Feynman diagrams for $B_{n0}$ in the NL$\chi$QM. We assign
  the initial and final pion momenta as $p_{i}$ and $p_{f}$,
  respectively, while the momentum transfer as $q$. We also define the
  loop momenta as $k_{a}=k-\frac{p_{i}}{2}-\frac{q}{2}$,
  $k_{b}=k+\frac{p_{i}}{2}-\frac{q}{2}$, and
  $k_{c}=k+\frac{p_{i}}{2}+\frac{q}{2}$.}        
\label{FIG0}
\end{figure}

The three-point correlation function in Eq.~(\ref{eq:TENSOR}) can be
easily calculated by a functional differentiation with respect to the
pion and external tensor fields, which leads to the following two
terms for the $B_{10}(Q^2)$: 
\begin{equation}
\label{eq:MAT1}
\frac{\delta ^{3}\mathcal{S}_{\mathrm{eff}}[m,\pi,T]}
{\delta T\, \delta \pi^{a} \delta \pi^{b}}
\Big|_{T=0}=
-\frac{1}{F^{2}_{\pi}}\mathrm{Sp}
\left[
\frac{1}{i\rlap{\,/}{D}}\sqrt{M}\gamma_{5}\tau^{a}\sqrt{M}
\frac{1}{i\rlap{\,/}{D}}\sqrt{M}\gamma_{5}\tau^{b}\sqrt{M}
\frac{1}{i\rlap{\,/}{D}}\sigma_{\mu\nu}
\right]
-\frac{i}{2F^{2}_{\pi}}\mathrm{Sp}
\left[
\frac{1}{i\rlap{\,/}{D}}\sqrt{M}\tau^{a}\tau^{b}\sqrt{M}
\frac{1}{i\rlap{\,/}{D}}\sigma_{\mu\nu}
\right],
\end{equation}
where we have introduced the shorthand notations $M=M(i\partial)$ 
and  $i\rlap{\,/}{D}=i\rlap{/}{\partial}+im+iM(i\partial) = 
i\rlap{/}{\partial}+i\bar{M}(i\partial)$. The trace
over the isospin space yields
$\mathrm{tr}_{\tau}[\tau^{a}\tau^{b}]=2\delta^{ab}$. One can also do
for the $B_{20}(Q^2)$ similarly. Having performed the functional trace
and the trace over the color space, we arrive at the matrix elements
for the $B_{(10,20)}(Q^2)$, corresponding to Eq.~(\ref{eq:TENSOR}), as 
follows:  
\begin{eqnarray}
\label{eq:MAT2}
\langle \pi^{+}(p_{f})|q^{\dagger}(0)\sigma_{ab}q(0)
|\pi^{+}(p_{i})\rangle&=& \underbrace{-\frac{2N_{c}}{F^{2}_{\pi}}
\int\frac{d^4k}{(2\pi)^4}
\mathrm{Tr}_{\gamma}
\left[
\frac{1}{i\rlap{\,/}{D}_{a}}\sqrt{M_{a}}\gamma_{5}\sqrt{M_{b}}
\frac{1}{i\rlap{\,/}{D}_{b}}\sqrt{M_{b}}\gamma_{5}\sqrt{M_{c}}
\frac{1}{i\rlap{\,/}{D}_{c}}\sigma_{ab}
\right]}_{\mathrm{(A)}}
\cr
&&\underbrace{
-\frac{iN_{c}}{F^{2}_{\pi}}\int\frac{d^4k}{(2\pi)^4}
\mathrm{Tr}_{\gamma}\left[
\frac{1}{i\rlap{\,/}{D}_{b}}\sqrt{M_{b}}
\frac{1}{i\rlap{\,/}{D}_{c}}\sqrt{M_{c}}\sigma_{ab}
\right]}_{\mathrm{(B)}},
\cr
\langle \pi^{+}(p_{f})|q^{\dagger}(0)\sigma_{ab}(i\tensor{D}\cdot a)q(0)
|\pi^{+}(p_{i})\rangle&=& \underbrace{-\frac{2N_{c}}{F^{2}_{\pi}}
\int\frac{d^4k}{(2\pi)^4}
\mathrm{Tr}_{\gamma}
\left[
\frac{1}{i\rlap{\,/}{D}_{a}}\sqrt{M_{a}}\gamma_{5}\sqrt{M_{b}}
\frac{1}{i\rlap{\,/}{D}_{b}}\sqrt{M_{b}}\gamma_{5}\sqrt{M_{c}}
\frac{1}{i\rlap{\,/}{D}_{c}}\sigma_{ab}[(k+\frac{p_i}{2})\cdot a]
\right]}_{\mathrm{(A)}}
\cr
&&\underbrace{
-\frac{iN_{c}}{F^{2}_{\pi}}\int\frac{d^4k}{(2\pi)^4}
\mathrm{Tr}_{\gamma}\left[
\frac{1}{i\rlap{\,/}{D}_{b}}\sqrt{M_{b}}
\frac{1}{i\rlap{\,/}{D}_{c}}\sqrt{M_{c}}\sigma_{ab}[(k+\frac{p_i}{2})\cdot
a] \right]}_{\mathrm{(B)}}.
\end{eqnarray}
The corresponding Feynman diagrams to the two terms (A) and (B) in the
right-hand side of Eq.~(\ref{eq:MAT2}) are depicted in
Fig.~\ref{FIG0}, respectively. 
The relevant momenta are also defined as:    
\begin{eqnarray}
\label{eq:MOM}
k_{a}&=&k-\frac{p_{i}}{2}-\frac{q}{2},
\,\,\,\,
k_{b}=k+\frac{p_{i}}{2}-\frac{q}{2},
\,\,\,\,
k_{c}=k+\frac{p_{i}}{2}+\frac{q}{2}.
\end{eqnarray}
In order to evaluate the matrix element, we define the initial and
final pion momenta in the Breit (brick-wall) frame in Euclidean space
as done in Ref.~\cite{Nam:2007gf}: 
\begin{equation}
\label{eq:MOM2}
p_{i}=\left(-\frac{Q}{2},0,0,i\sqrt{\frac{Q^{2}}{4}+m^{2}_{\pi}} \right),
\,\,\,\,
p_{f}=\left(\frac{Q}{2},0,0,i\sqrt{\frac{Q^{2}}{4}+m^{2}_{\pi}} \right),
\,\,\,\,
q=\left(Q,0,0,0 \right).
\end{equation}
We also have chosen the auxiliary vectors for definiteness as
$a=(0,1,0,i)$ and $b=(1,0,1,0)$, which satisfy the conditions
mentioned in Section II. The denominators become
$\rlap{\,/}{D}_{a,b,c}=\rlap{/}{k}_{a,b,c}+i\bar{M}_{a,b,c}$ in
Eq.~(\ref{eq:MAT2}). The momentum-dependent effective quark mass
$M_{a,b,c}$ can be also defined by using Eqs.~(\ref{eq:EFFM}) and 
(\ref{eq:MOM}). 
\section{Numerical results and Discussions}
\begin{figure}[h]
\begin{tabular}{cc}
\includegraphics[width=8.5cm]{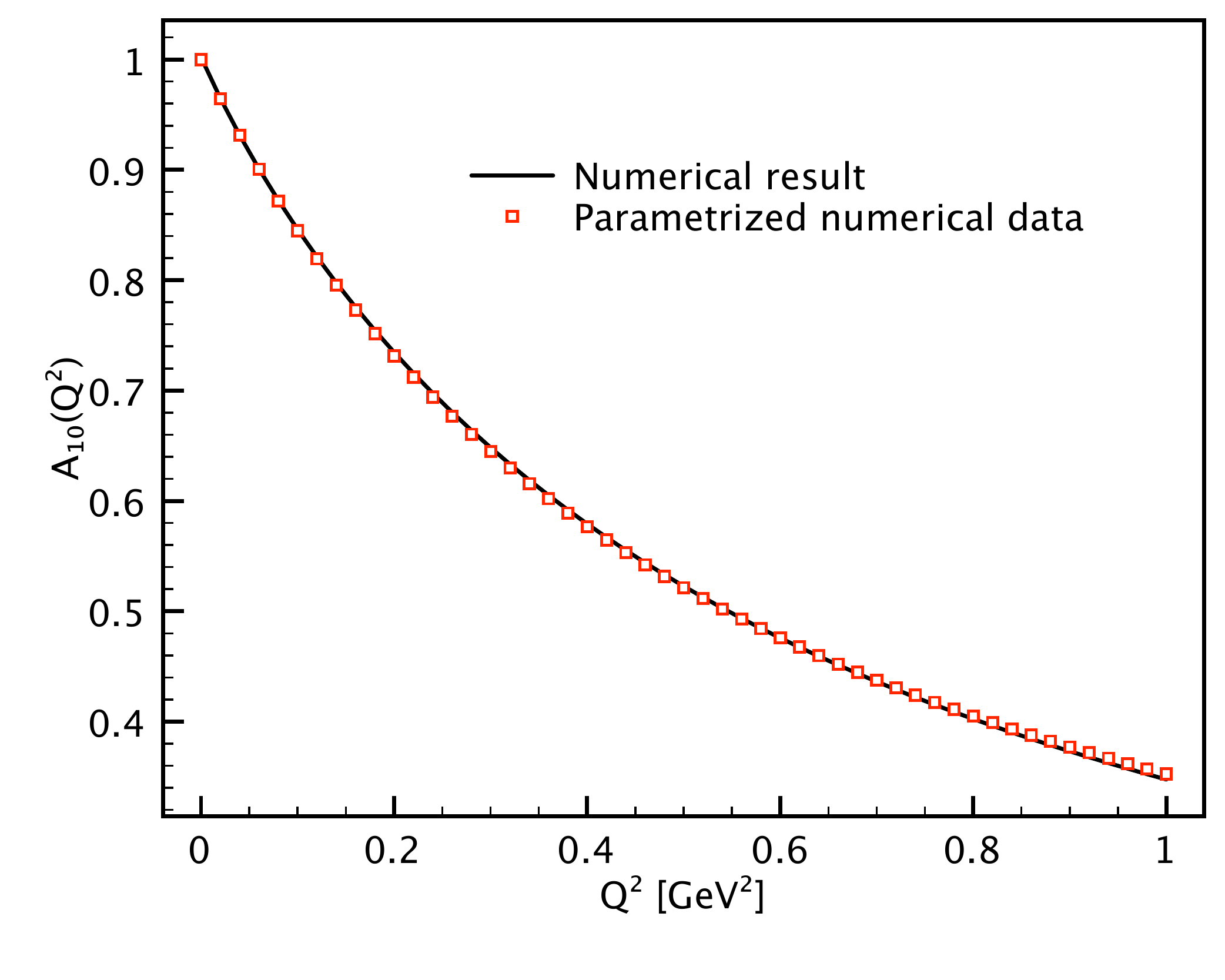}
\includegraphics[width=8.5cm]{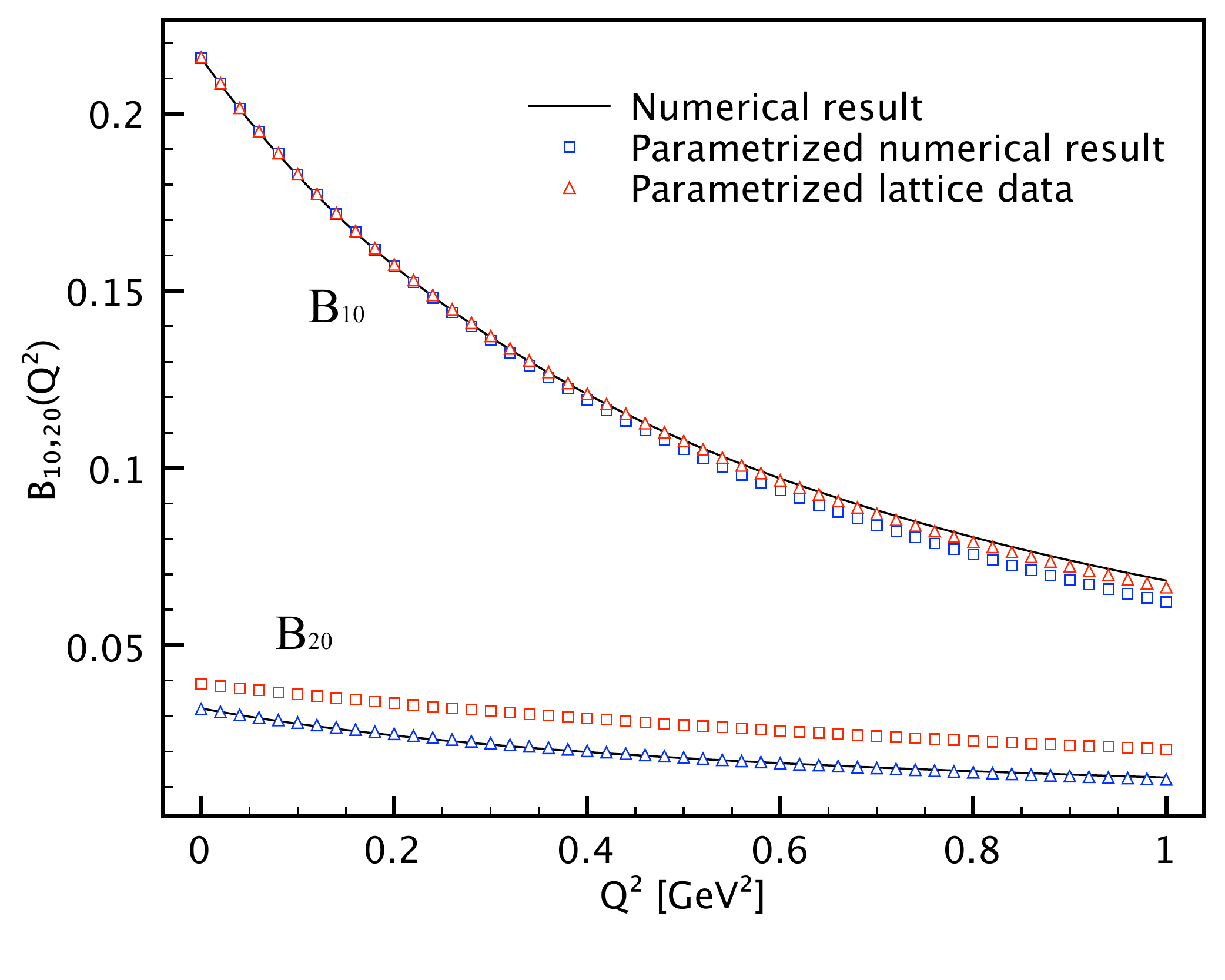}
\end{tabular}
\caption{The numerical results for the electromagnetic form factor
$A_{10}(Q^{2})$ are presented in the left panel and those for the
tensor form factors $B_{10}(Q^{2})$  and $B_{20}(Q^{2})$ in the
right panel. The solid curves depict the numerical results, whereas
their parametrizations 
given respectively in Eqs.~(\ref{eq:monopol}) and (\ref{eq:FFS}) are
denoted by the squares. We also show 
the parametrized lattice data~\cite{Brommel:2007xd} for the tensor
form factors, designated by the triangles in the right panel.}          
\label{FIG1}
\end{figure}

We first discuss the numerical results of the tensor form factors of the
pion. Figure~\ref{FIG1} draws the numerical results for the
electromagnetic form factor of the pion $A_{10}$ in the left panel and
its tensor form factor $B_{10}$ in the right panel 
as functions of $Q^{2}$ in the range of $0\le  
Q^{2}\le1\,\mathrm{GeV}^{2}$. 
However, we want to mention that there is a caveat. Since we need the
results of the form factors in principle up to infinite $Q^2$ in order
to perform the Fourier transform given in Eq.~(\ref{eq:FT}), 
we will use the parametrized one, as we will discuss later in the
context of the lattice data.

The numerical results for $A_{10}(Q^{2})$ are taken from
Ref.~\cite{Nam:2007gf}. Though we have already discussed those for
the electromagnetic form factor in detail in Ref.~\cite{Nam:2007gf},
we want to recapitulate them in the context of the lattice data. It is
well known that the electromagnetic form factor can be parametrized by
a monopole form  
\begin{equation}
\label{eq:monopol}
 F_\pi(Q^2)=A_{10}(Q^2)=\frac1{1+Q^2/M^2}. 
\end{equation}
The monopole mass $M$ was known to be 
$M=(0.714\pm0.004)$ GeV, based on the experimental
data~\cite{Amendolia:1986wj,Tadevosyan:2007yd,Horn:2006tm}.    
On the other hand, the lattice calculation yields
$M=(0.727\pm0.016)$ GeV with a linear chiral 
extrapolation to the physical pion mass taken into
account~\cite{Brommel:2006ww}.  The present result leads to
$M=0.738\,\mathrm{GeV}$, which indicates that that of the pion 
electromagnetic form factor is in good agreement with the lattice
data. We also obtain the squared charge radius of the pion $\langle
r^2\rangle = 0.456\,\mathrm{fm}^2$, while in the lattice QCD it was
evaluated to be $\langle r^2\rangle = (0.441\pm0.019)\,\mathrm{fm}^2$.
Considering the uncertainty of the lattice data, the present result is
in remarkable agreement with them.  In the left panel of
Fig.~\ref{FIG1}, we show the numerical result (solid
curve)~\cite{Nam:2007gf} and its monopole parametrization (sqaure) of
$A_{10}(Q^{2})$, using the values mentioned above and
Eq.~(\ref{eq:monopol}).  

In the right panel of Fig.~\ref{FIG1}, we draw the numerical results
for the tensor form factors $B_{10}(Q^{2})$ and $B_{20}(Q^{2})$ (solid
curve). In order to compare the present
results with the lattice data, it is crucial to consider the
evolution of the
scale~\cite{Barone:2001sp,Broniowski:2009zh,Broniowski:2010nt}, since 
the tensor current is not the conserved one. The tensor form factor is
evolved at the leading order (LO) by the following
equation~\cite{Barone:2001sp,Broniowski:2010nt}  
\begin{equation}
\label{eq:evolve}
B_{n0} (Q^2,\mu)=B_{n0}(Q^2,\mu_0) 
\left[\frac{\alpha(\mu)}{\alpha(\mu_0)}\right]^{{\gamma_{n}}/{(2\beta_0)}},      
\end{equation}
where we have used the anomalous dimensions $\gamma_1=8/3$ and
$\gamma_{2} =8$, and $\beta_0= 11N_c/3 - 2N_f/3$ ($N_c=3$ and $N_f=2$
in the present case).  Thus, the powers in the LO evolution equation
are given as $4/29$ and $12/29$ respectively for $n=1$ and $n=2$,
which indicate that the dependence of the tensor 
charge on the normalization point turns out to be rather weak. Note
that the anomalous dimension is simply the  same as that for the
nucleon tensor charge~\cite{Kim:1995bq}. We also take
$\Lambda_{\mathrm{QCD}}=0.248\,\mathrm{GeV}$  which was    
also used in evolving the nucleon tensor charges and tensor anomalous 
magnetic  moments~\cite{Ledwig:2010tu,Ledwig:2010zq}.  Since the
normalization point of the present model is around 
$0.6\,\mathrm{GeV}$, while the lattice calculation was carried out at
$\mu=2\,\mathrm{GeV}$, the scale factors turn out to be 
\begin{equation}
\label{eq:scale}
B_{10} (Q^2,\mu=2\,\mathrm{GeV})
=0.89\,B_{10} (Q^2,\mu_0=0.6\,\mathrm{GeV}) ,
\;\;\;\;
B_{20} (Q^2,\mu=2\,\mathrm{GeV})
=0.70\,B_{20}(Q^2,\mu_0=0.6\,\mathrm{GeV}). 
\end{equation}
Considering these scalings, we obtain
$B_{10}(0)=0.216$ and $B_{20}(0)=0.032$. In the lattice
calculation~\cite{Brommel:2007xd}, the tensor charge of the pion for
$n=1$ with the linear chiral extrapolation to the physical pion mass
in $m_\pi^2$ was estimated to be about $B_{10}(0)=0.216$, which is
almost identical to the present result. As for $n=2$, the
lattice data estimated about $B_{20}(0)=0.039$, which is about $20\,\%$
larger than the present one, but is still comparable. We want to
mention that one could use larger current-quark masses in order to
compare directly with the lattice data as done in
Ref.~\cite{Broniowski:2010nt}. However, it is rather unreliable in the 
present framework: Firstly it is nontrivial to include the larger
current quark 
mass~\cite{Musakhanov:1998wp,Musakhanov:2002vu,Musakhanov:2002xa}.  
Secondly, the present scheme is conceptually only valid when the
current quark mass is small, at least up to the strange current quark
mass. Thus, the NL$\chi$QM from the instanton vacuum
is a rather restricted one, so that we will compare the present
results with those of the lattice QCD with chiral extrapolation.  

\begin{figure}[t]
\begin{tabular}{cc}
\includegraphics[width=8.5cm]{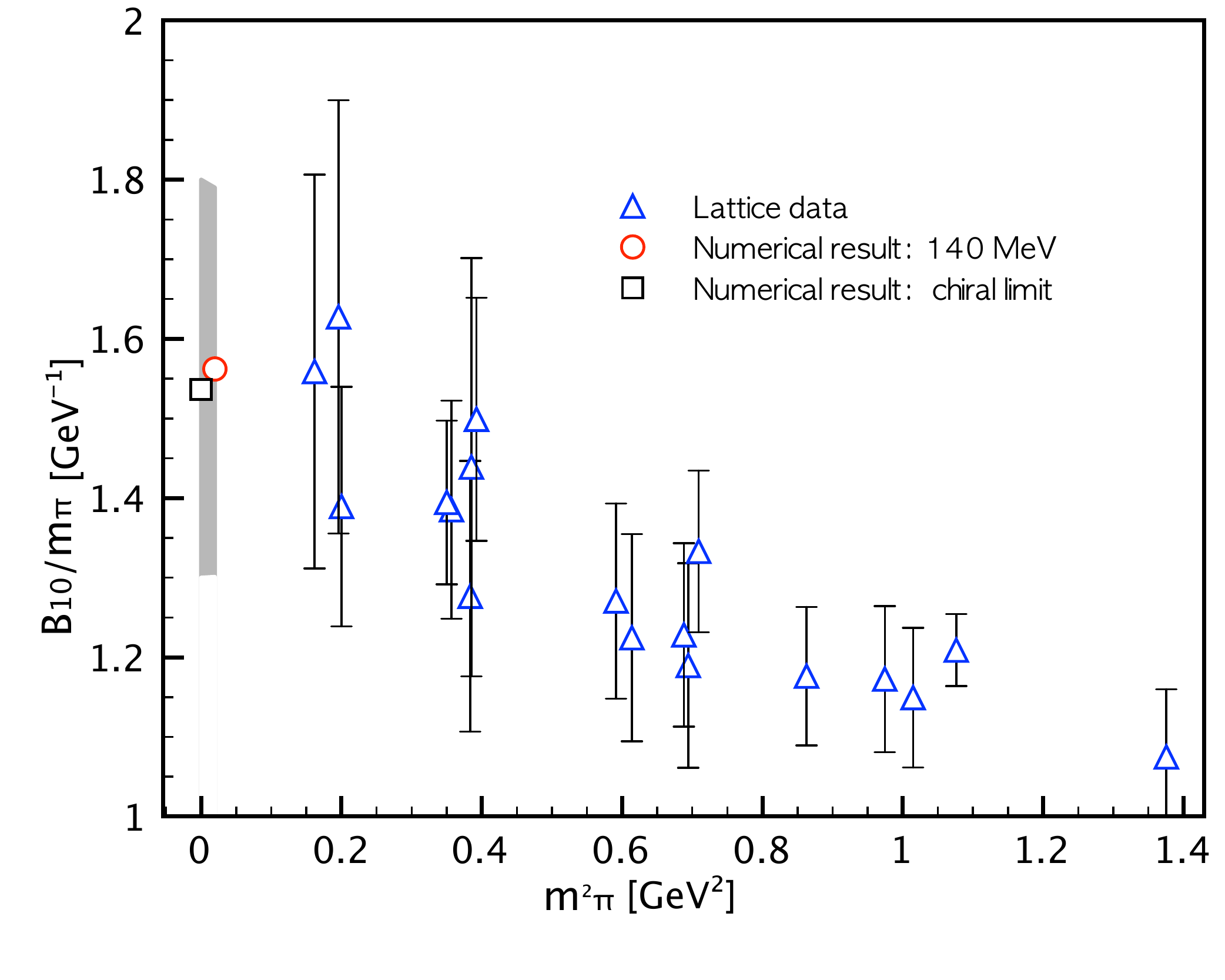}
\includegraphics[width=8.5cm]{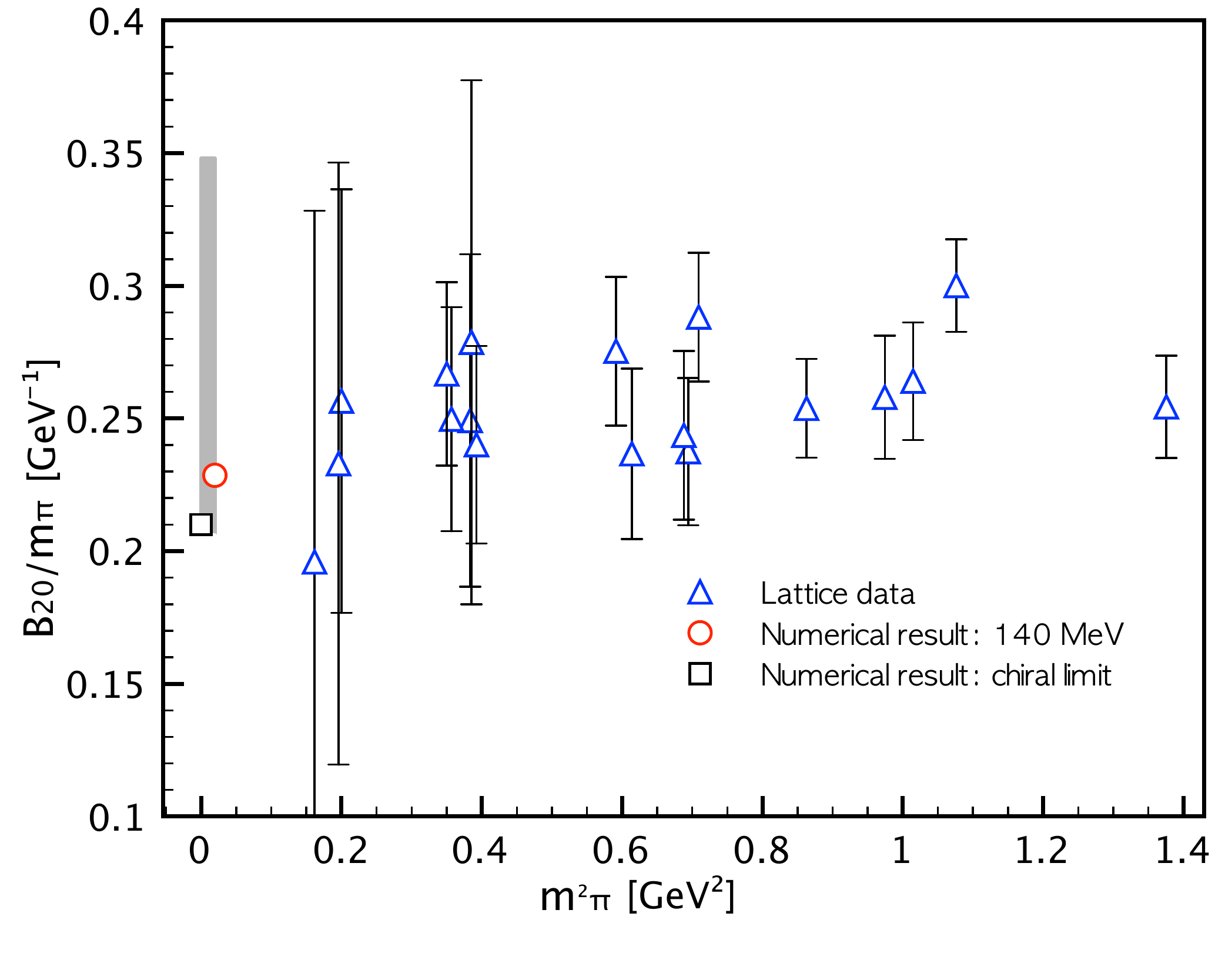}
\end{tabular}
\caption{$B_{10}(0)/m_{\pi}$ (left) and $B_{20}(0)/m_{\pi}$ (right) as
  functions of $m^{2}_{\pi}$. The numerical results for $m_{\pi}=0$ and
  $140$ MeV are given by the square and circle. The lattice data
  (triangle) are taken from Ref.~\cite{Brommel:2007xd}. The shaded
  areas for $m_{\pi}=(0\sim140)$ MeV represent the lattice fits.}        
\label{FIG10}
\end{figure}

In Fig.~\ref{FIG10}, we present the $m_\pi$ dependence of the pion
tensor form factor scaled by the pion mass as a function of
$m^{2}_{\pi}$ for $m_{\pi}=0$ (square) and $140$ MeV (circle). As
shown in Fig.~\ref{FIG10}, the result in the chiral limit is slightly
smaller than that with $m_\pi = 140\,\mathrm{MeV}$. As for the case
with $m_{\pi}=0$, we take the current quark mass $m=0$. The shaded
bands represent the fits from the lattice
calculation~\cite{Brommel:2007xd}. 

A simple $p$-pole parametrization of GFFs was used in
Ref.~\cite{Brommel:2007xd} to get the tensor form factor of the pion:     
\begin{equation}
\label{eq:PARA}
B_{n0}(Q^{2})=B_{n0}(0)
\left[1+\frac{Q^{2}}{p_n\,m^{2}_{p_n}} \right]^{-p_n}.
\end{equation}
In this parametrization, the lattice QCD simulation estimated the 
pole mass $m_{p_1}=(0.756\pm0.095)$ GeV and $m_{p_2}=(1.130\pm0.265)$
at $m_{\pi}=140$ MeV with chiral extrapolation. Considering the
condition $p>1.5$ for the regular behavior of the probability density
at $b_{\perp}\to0$~\cite{Diehl:2005jf} and following
Ref.~\cite{Brommel:2007xd}, we take $p_{1}=p_{2}=1.6$ as a trial. If
this is the case, Eq.~(\ref{eq:PARA}) gives us $m_{p_1}\approx0.761$
GeV and $m_{p_2}\approx0.864$ GeV to reproduce the present results,
which is compatible with that of the lattice simulation. Taking into
account these results, we can write the $p$-pole parametrized tensor
form factor as follows:  
\begin{equation}
\label{eq:FFS}
B_{10}(Q^{2})=0.216
\left[1+\frac{Q^{2}}{1.6\times0.761^{2}\,\mathrm{GeV}^{2}}
\right]^{-1.6},\,\,\,\, 
B_{20}(Q^{2})=0.032
\left[1+\frac{Q^{2}}{1.6\times0.864^{2}\,\mathrm{GeV}^{2}}
\right]^{-1.6}, 
\end{equation}
The result of this parametrized one in Eq.~(\ref{eq:FFS}) is also
depicted in the right panel of Fig.~\ref{FIG1} (square), and reproduces 
well the present numerical one. We also compare our result with the
lattice one in the right panel of Fig.~\ref{FIG1}. As in the case of
the electromagnetic form factors, the present results are in excellent
agreement with the lattice data (triangle). Note that in the present
framework we do not have any adjustable free parameter.  

\begin{figure}[h]
\begin{tabular}{cc}
\includegraphics[width=7.5cm]{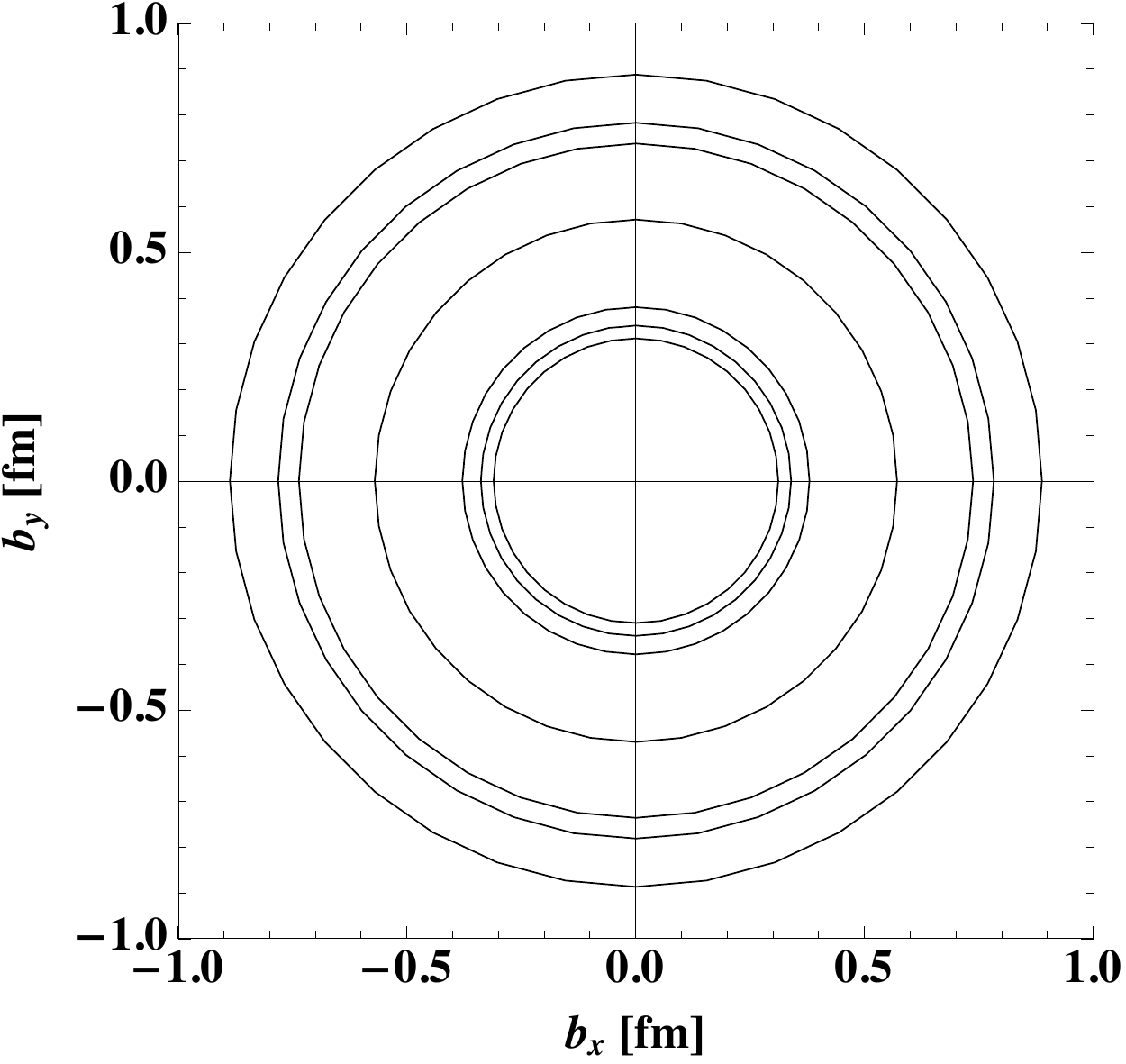}
\includegraphics[width=7.5cm]{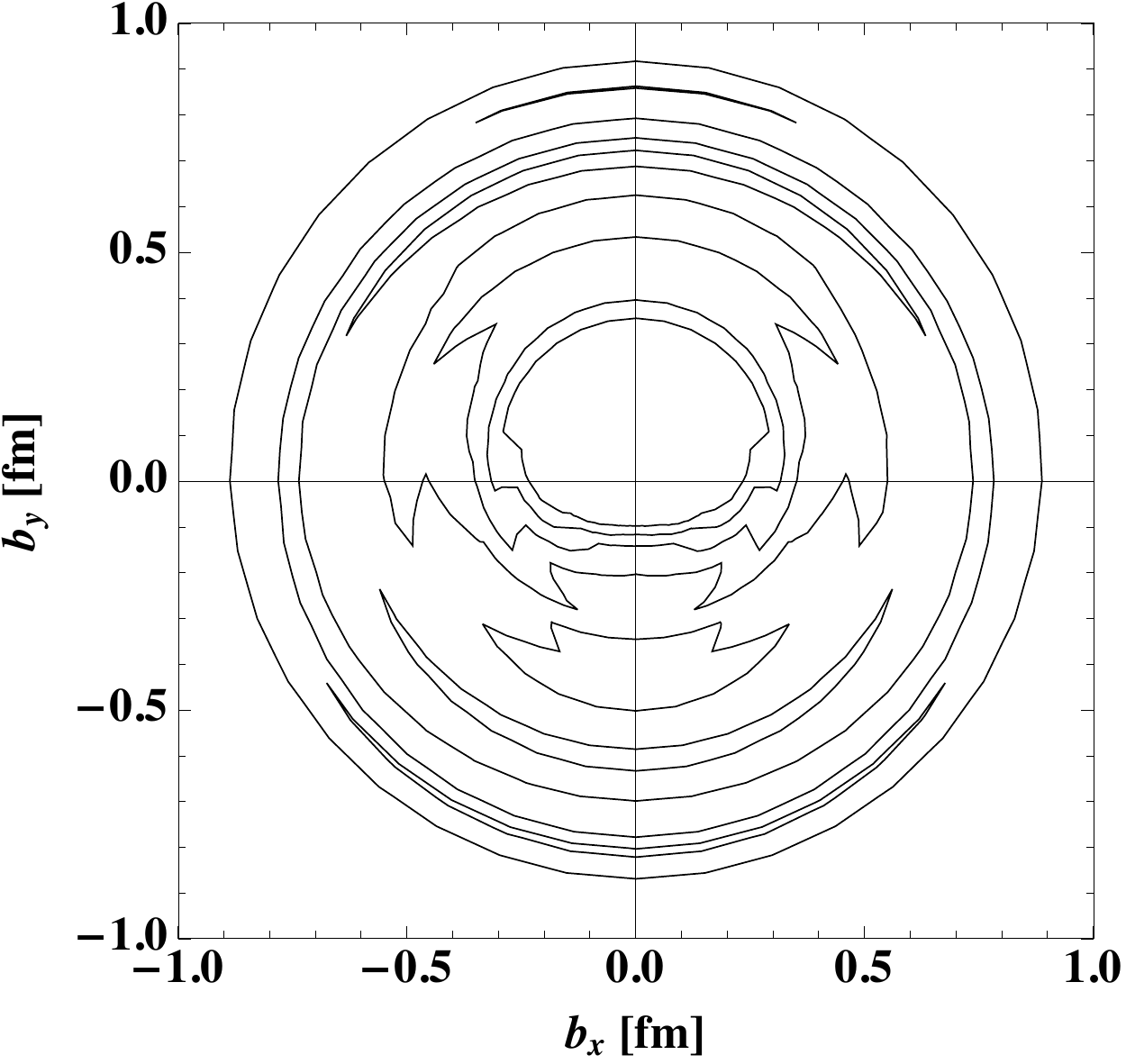}
\end{tabular}
\begin{tabular}{cc}
\includegraphics[width=8.5cm]{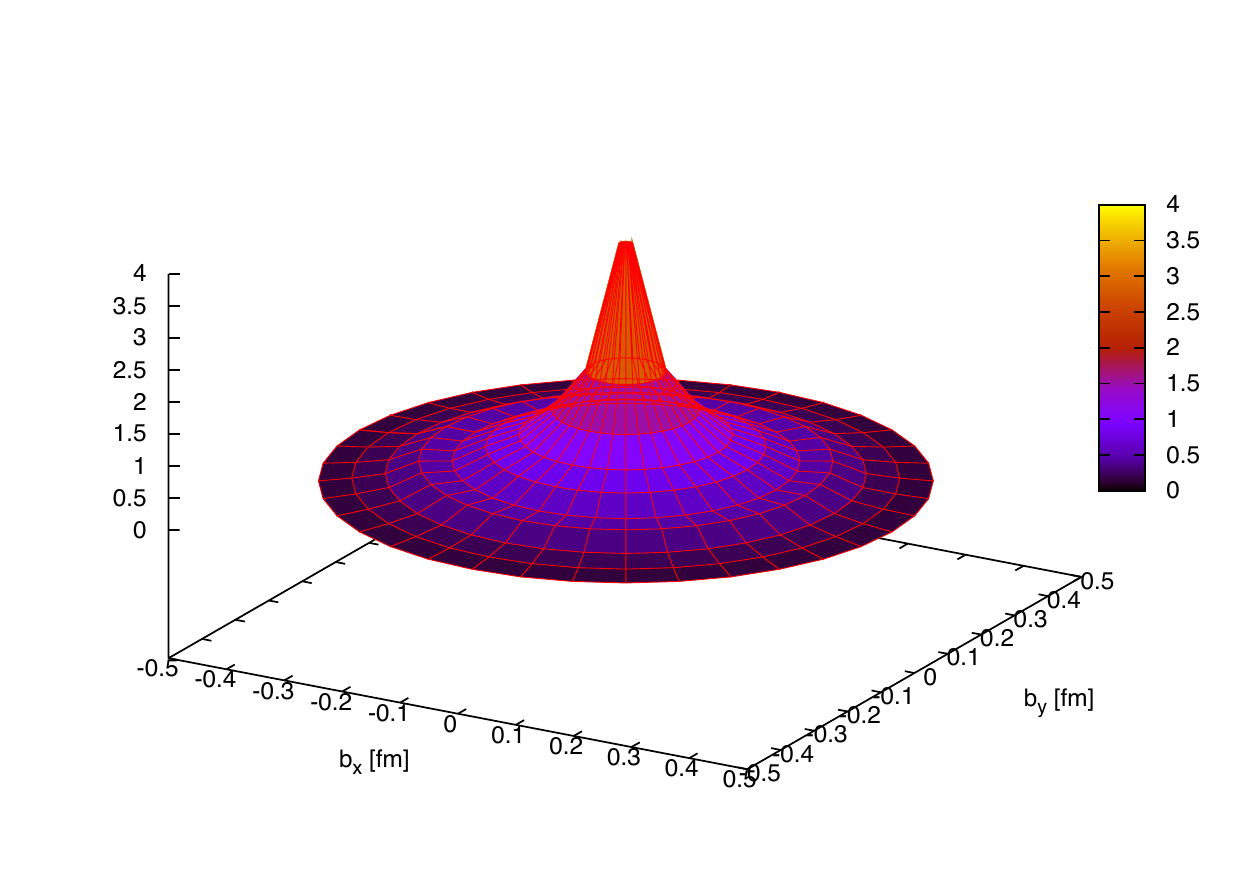}
\includegraphics[width=8.5cm]{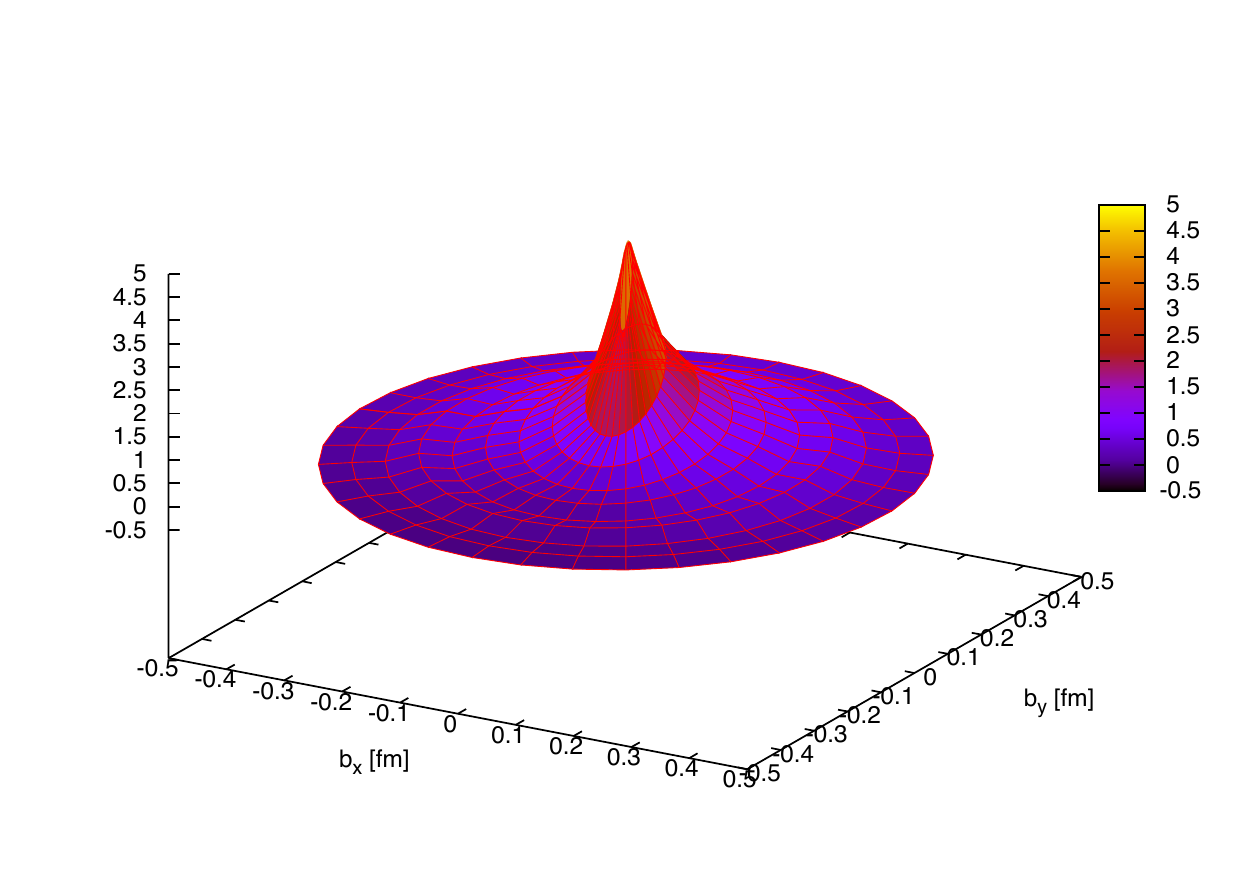}
\end{tabular}
\caption{In the upper panels, we show unpolarized (left) and polarized
  (right) probability densities, $\rho_{1}$ in Eq.~(\ref{eq:DENSITY})
  as a function of   $b_{x}$ and $b_{y}$ for
  $s_{\perp}=(+1,0)$. Similarly, we show their three dimensional
  profiles in the lower panels.}         
\label{FIG2}
\end{figure}

We are now in a position to discuss the results of the probability
densities of the transversely polarized quarks inside the pion,
defined in Eq.~(\ref{eq:DENSITY}). In the upper-left panel of
Fig.~\ref{FIG2}, we show the unpolarized probability density with the
tensor form factor turned off.  As expected, the quarks are
distributed symmetrically on the $b_x$-$b_y$ plane. On the other hand,
if we switch on the tensor form factor, the spatial distribution of a
transversely polarized quark inside the pion ($\pi^+$) gets
distorted as shown in the upper-right panel of Fig.~\ref{FIG2}.  
Its maximum value is also shifted to the $b_y$ direction in comparison
to that for the unpolarized one. Thus, it is interesting to examine
the average transverse shift which is defined
as~\cite{Brommel:2007xd}:   
\begin{equation}
\label{eq:SHIFT}
\langle b_{y}\rangle=\frac{\int
  d^2\,b_\perp\,b_{y}\,\rho(b_\perp,s_\perp)}{\int
  d^2\,b_\perp\,\rho(b_\perp,s_\perp)}  
=\frac{1}{2m_{\pi}}\frac{B_{10}(0)}{A_{10}(0)},
\end{equation}
where we have chosen $s_{\perp}=(+1,0)$. Using Eq.~(\ref{eq:FFS}) and
$m_{\pi}=140$ MeV, we obtain $\langle b_{y}\rangle=0.152$ fm, which is
almost the same as that of the lattice calculation $\langle
b_{y}\rangle=(0.151\pm0.024)$ fm. This finite value of $\langle
b_{y}\rangle $ measures how much the polarized probability density is
distorted in the transverse plane. If we take the spin quantized along
the $y$ axis, i.e. $s_{\perp}=(0,+1)$, the result of the polarized
probability density is similar but rotated by $90^{\circ}$
clockwise. We also note that the present results are almost equivalent
to those given by the lattice simulation~\cite{Brommel:2007xd}. In the
lower panel of Fig.~\ref{FIG2}, we show the three-dimensional profiles
for the unpolarized (left) and transversely polarized (right)
distributions as functions of $b_{x}$ and $b_{y}$. One can obviously
see that the maximum of the transversely polarized probability density
is shifted and distorted. 

In Fig.~\ref{FIG3}, we draw the probability density as a function of
$b_{y}$ at $b_{x}\approx0.2$ fm, comparing it with that of the
lattice calculation. As expected, the present result is almost
identical to that of the lattice QCD. We summarize the main results of
the present work in Table~\ref{TABLE0}:  

\begin{figure}[t]
\includegraphics[width=8.5cm]{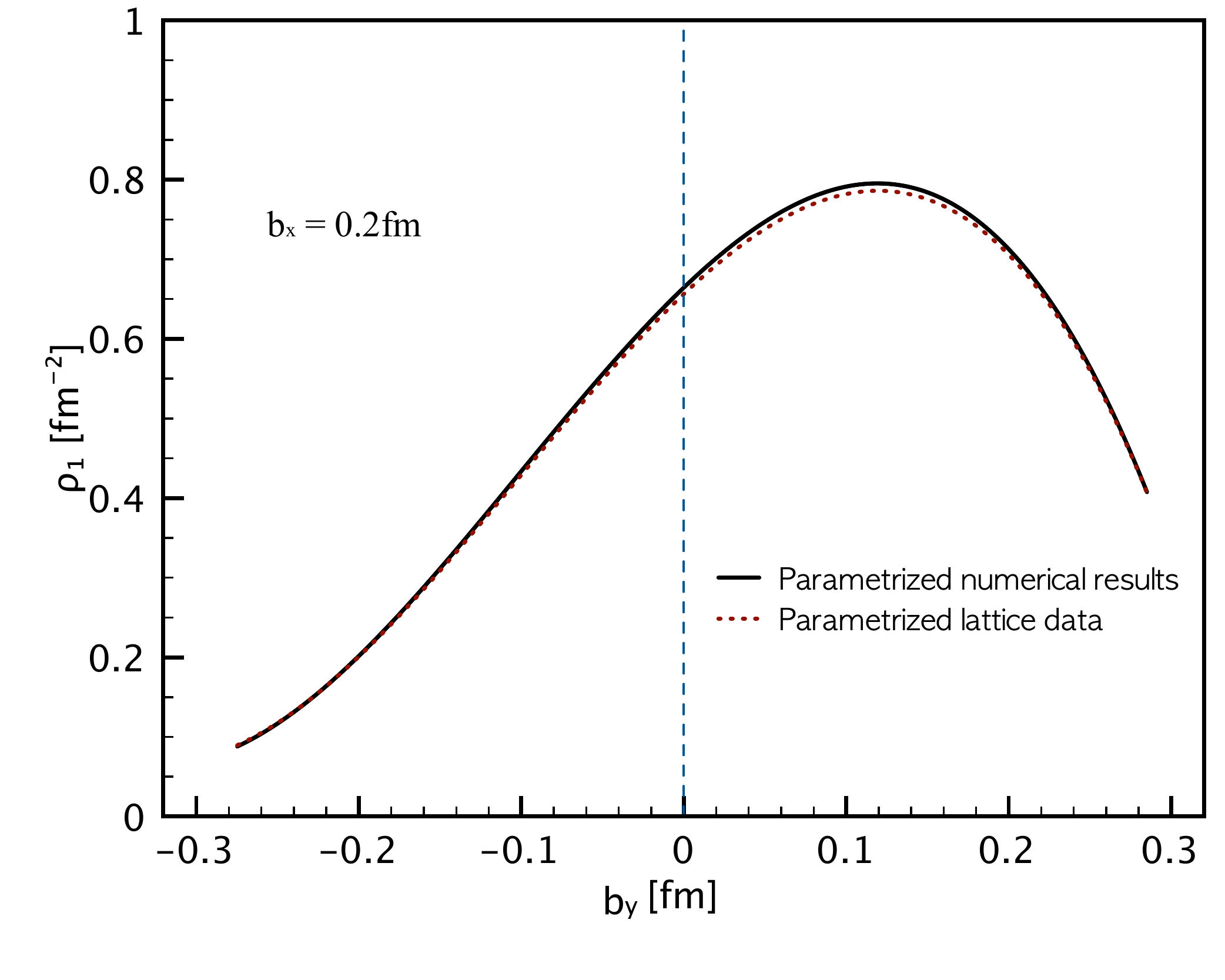}
\caption{Comparison between the polarized probability densities from
  the present (square) and lattice (triangle) results for
  $b_{x}\approx0.2$ fm.}         
\label{FIG3}
\end{figure}

\begin{table}[b]
\begin{tabular}{l|ccc|ccc}
\hline\hline
$m_{\pi}=140$ MeV&$B_{10}(0)$&$m_{p_1}$ [GeV]&$\langle b_{y}\rangle$ [fm]
&$B_{20}(0)$&$m_{p_2}$ [GeV]\\
\hline
Present work&$0.216$&$0.762$&$0.152$&$0.032$&$0.864$\\
\hline
Lattice QCD~\cite{Brommel:2007xd}&$0.216\pm0.034$ & $0.756\pm0.095$ &
$0.151$&$0.039\pm0.099$ &$1.130\pm0.265$\\ 
\hline\hline
\end{tabular}
\caption{Numerical results for $B_{n0}(0)$, $m_{p_n}$, and $\langle
  b_{y}\rangle $ in comparison with those of the lattice
  simulation~\cite{Brommel:2007xd} with the chiral extrapolation to
  the physical pion mass $m_{\pi}=140$ MeV.} 
\label{TABLE0}
\end{table}

\section{Summary and conclusion}
In the present work, we have aimed at investigating the tensor form
factors of the pion, $B_{10}$ and $B_{20}$, using the nonlocal chiral
quark model from the instanton vacuum. Combining it with the
electromagnetic form factor of the pion~\cite{Nam:2007gf}, we were
able to evaluate the transversely polarized density of quarks inside
the pion as functions of $(b_{x},b_{y})$ in comparison with the
lattice simulation~\cite{Brommel:2007xd}. 

We first recapitulated the electromagnetic form factor of the pion
computed previously in the context of the lattice calculation. 
We found that the monopole mass is $M=0.738$ GeV which is in a good
agreement with that from the lattice QCD $M=0.727$ GeV as well as the
experimental data $M=(0.714\pm0.004)$ GeV. It indicates that the $Q^2$
dependence of the electromagnetic form factor is well reproduced
within the present work and compatible with the lattice results. We
also presented the squared charge radius of the pion $\langle r^2
\rangle = 0.456\,\mathrm{fm}^2$ which is again in very good agreement  
with the lattice result $\langle r^2 \rangle =
(0.441\pm0.019)\,\mathrm{fm}^2$.  

We calculated the tensor form factor of the pion within the same
framework as done in previous works. In order to compare
the results with the lattice data, we evolved them from the
$\mu=\bar{\rho}^{-1} = 600\,\mathrm{MeV}$ to the scale at
which the lattice calculation was performed
($\mu=2\,\mathrm{GeV}$). We also carried out the $p$-pole
parametrization as in the lattice QCD.  The results for the tensor
form factor were obtained as follows: The tensor form factors of the
pion at $Q^2=0$ $(B_{(10)}(0),\,B_{(20)}(0))=(0.216,\,0.032)$ and the pole
mass $m_{p_{(1,2)}}=(0.762,0.864)\,\mathrm{GeV}$. Being compared to
the lattice results with chiral extrapolation,
i.e. $B_{(10,20)}(0)=(0.216\pm0.034,0.039\pm0.099)$ and
$m_{p_{(1,2)}}=(0.756\pm 0.095,1.130\pm0.265)$ GeV, they were found to
be almost identical and comparable to those of the lattice QCD. In
particular, these results are remarkable, considering the fact that
the present scheme does not contain any adjustable parameter. 

Having combined the results of the tensor form factor with those of
the electromagnetic one, we obtained straightforwardly the
probability density of transversely polarized quarks inside the
pion. It turned out that the spatial distribution of the quarks on the
transverse plane were distorted, compared to that of the unpolarized
quarks. Moreover, the maximum value of the density is shifted to the
$b_y$ direction. In order to examine this shift, we also calculated
the average value of $b_y$ which turned out to be $\langle b_y\rangle
= 0.152\,\mathrm{fm}$. It is in an excellent agreement with the
lattice result $\langle b_y\rangle = 0.151\,\mathrm{fm}$.

Finally, It is worth mentioning that it is also of great importance to
study the spin structure of the kaon, since it sheds light on the role
of flavor SU(3) symmetry breaking inside the kaon. Related works are
under progress and will appear elsewhere. 
\section*{Acknowledgments}
The authors are grateful to Ph.~H\"agler (the QCDSF/UKQCD
collaborations) for providing us with the data from the lattice
calculation. S.i.N. is thankful to the hospitality during his visiting
Inha University, where this work was performed. The work  of 
H.Ch.K. was supported by Basic Science Research Program through the
National Research Foundation of Korea (NRF) funded by the Ministry of 
Education, Science and Technology (grant number: 2010-0016265). The
work of S.i.N. was supported by the grant NRF-2010-0013279 from
National Research Foundation (NRF) of Korea. The numerical
calculations were partially performed on SAHO at RCNP, Osaka
University.  

\end{document}